\documentclass[onecolumn]{aastex}

\usepackage{graphicx}
\usepackage{dcolumn}
\usepackage{bm}
\usepackage{color}
\usepackage{amsfonts}
\usepackage{amssymb}
\usepackage{amsmath}

\input epsf

\begin{document}

\title{\Large{Observational Study of Higher Dimensional Magnetic Universe\\ in Non-linear Electrodynamics}}

\author{{\bf Chayan Ranjit}\altaffilmark{1}}
\author{{\bf Shuvendu Chakraborty}\altaffilmark{2}}
\author{{\bf Ujjal Debnath}\altaffilmark{3}}

\altaffiltext{1}{{Department of Mathematics, Seacom Engineering
College, Howrah - 711 302, India. Email: chayanranjit@gmail.com}}

\altaffiltext{2}{{Department of Mathematics, Seacom Engineering
College, Howrah - 711 302, India.
Email:shuvendu.chakraborty@gmail.com}}

\altaffiltext{3}{{Department of Mathematics, Bengal Engineering
and Science University, Shibpur, Howrah-711 103, India. Email:
ujjaldebnath@yahoo.com}}

\date{\today}

\begin{abstract}
In this work, we have considered the flat FRW model of the
universe in $(n+2)$-dimensions filled with the dark matter and the
magnetic field. We present the Hubble parameter in terms of the
observable parameters $\Omega_{m0}$ and $H_{0}$ with the redshift
$z$ and the other parameters like $B_{0},\omega,
\mu_{0},\delta,n,w_{m}$. The natures of magnetic field $B$,
deceleration parameter $q$ and $Om$ diagnostic have also been
analyzed for accelerating expansion of the universe. From Stern
data set (12 points), we have obtained the bounds of the arbitrary
parameters by minimizing the $\chi^{2}$ test. The best-fit values
of the parameters are obtained by 66\%, 90\% and 99\% confidence
levels. Now to find the bounds of the parameters ($B_{0},\omega$)
and to draw the statistical confidence contour, we fixed four
parameters $\mu_{0},\delta,n,w_{m}$. Here the parameter $n$
determines the higher dimensions and we perform comparative study
between three cases : 4D $(n=2)$, 5D $(n=3)$ and 6D $(n=4)$
respectively. Next due to joint analysis with BAO observation, we
have also obtained the bounds of the parameters ($B_{0},\omega$)
by fixing other parameters $\mu_{0},\delta,n,w_{m}$ for 4D, 5D and
6D. The best fit of distance modulus for our theoretical model and
the Supernova Type Ia Union2 sample are drawn for different
dimensions.
\end{abstract}

\keywords{Higher Dimension; Om Diagnostic; Observational Data;
Observational Constraints.}

\maketitle

\section{\normalsize\bf{Introduction}}

The origin of the classical Einstein field equations are Maxwell's
electrodynamics which leads to the singular isotropic Friedmann
solutions. Over the last few years the standard cosmological
model, based on Friedmann-Robertson-Walker (FRW) geometry with
Maxwell's electrodynamics has got sufficient amount of interest
and many significant result are obtained \cite{Kolb1990, Murphy,
Sitter, Novello1979, Novello1993, Breton2005,Breton2000,
Novello2005, Klippert}. Recently the non-linear electrodynamics
(NLED) is a very interesting subject of research in general
relativity. In 1934, Born and Infield \cite{Born1934} wanted to
modify the standard Maxwell theory in order to eliminate the
problem of infinite energy of electron. In present time a new
approach \cite{Lorenci} has been taken to avoid the cosmic
singularity through a nonlinear extension of the Maxwell's
electromagnetic theory and black hole solution can be obtained
\cite{Kats2007,Anninos2009,Cai2008}. Another interesting feature
can be viewed that for construction of regular black hole
solutions \cite{Ayón-Beato1998,Ayón-Beato1999,Salazar1987},
nonlinear electrodynamics theories are most powerful tool. The
solution of the Einstein field equations together with NLED
signifies the nonlinear effects in strong gravitational and
magnetic fields. In the standard Maxwell Lagrangian, the nonlinear
terms can be added by imposing the existence of symmetries such as
parity conservation, gauge invariance, Lorentz invariance, etc.
\cite{Novello1996,G1996}, as well as by the introduction of first
order quantum corrections invariance to the Maxwell
electrodynamics \cite{Heisenberg,Schwinger}.\\

Our theoretical models are continuously testing by the different
observational data. Currently the universe is expanding with
acceleration which is verified by different observations of the
SNeIa \cite{Perlmutter1998,Perlmutter1999,Riess1998,Riess2004},
large scale redshift surveys  \cite{Bachall,Tedmark}, the
measurements of the cosmic microwave background (CMB)
\cite{Miller,Bennet} and WMAP \cite{Briddle,Spergel}. The
observational facts are not clearly described by the standard big
bang cosmology with perfect fluid. Recently several interesting
mechanisms such as loop quantum cosmology \cite{Asthekar},
modified gravity \cite{Cognola}, higher dimensional phenomena
\cite{Chakraborty2010,Ranjit}, Brans-Dicke theory \cite{Brans},
brane-world model \cite{Gergely} and so on, suggested that some
unknown matters are responsible for accelerating scenario of the
universe which has positive energy density and sufficient negative
pressure, known as dark energy \cite{Sahni2000,Paddy}. The most
suitable type of dark energy for that scenario is the scalar field
or quintessence \cite{Peebles} in which the potential dominates
over the kinetic term. In the present time several cosmological
models have been constructed by introducing dark energies such as
phantom \cite{Caldwell,Bronnikov,Chang}, tachyon scalar field
\cite{Sen,Balart,Farajollahi,del}, hessence \cite{Wei}, dilaton
scalar field \cite{Morris,Marcus}, K-essence scalar field
\cite{Armen,Bouhmadi,Malquarti}, DBI essence scalar field
\cite{Spalinski,Martin} and many others. Recent observational
evidence suggests that the present Universe is formed of $\sim$
26\% matter (baryonic + dark matter) and $\sim$ 74\% of a smooth
vacuum energy component and about 0.01\% of the thermal CMB
component. The information about the structure formation process
and other important cosmic observable are obtained by the angular
power spectrum of CMB components in anisotropic.\\

Brief review of Maxwell's electrodynamics and non-linear
electrodynamics are presented in section II. The basic equations
in $(n+2)$-dimensional FRW universe and their solutions are given
in section III for interacting model. The nature of Om-diagnostic
are studied also. The observational data analysis mechanism for
non-linear electrodynamic are described in section IV. The
$\chi^{2}_{min}$ test for best fit values of the observational
parameters are investigated with Stern and then Stern+BAO joint
data analysis. The best fit of distance modulus for our
theoretical model and the Supernova Type Ia Union2 sample are
drawn for different dimensions.
Finally, some observational conclusions are drawn in section V.\\

\section{\normalsize\bf{Non-linear Electrodynamics }}

The Lagrangian density in Maxwell's electrodynamics can be written
as \cite{Camara2004}

\begin{equation}
{\cal
L}_{(MAXWELL)}=-\frac{1}{4\mu_{0}}~F^{\mu\nu}F_{\mu\nu}=-\frac{1}{4\mu_{0}}~F
\end{equation}

where $F^{\mu\nu}$ is the electromagnetic field strength tensor,
$F$ is the electromagnetic field and $\mu_{0}$ is the magnetic
permeability. The canonical energy-momentum tensor is then given
by

\begin{equation}
T^{(MAXWELL)}_{\mu\nu}=\frac{1}{\mu_{0}}\left(F_{\mu\alpha}F^{\alpha}_{\nu}+\frac{1}{4}~Fg_{\mu\nu}
\right)
\end{equation}

The general class of Lagrangian for non-linear electromagnetic
field \cite{Novello2} can be written in the form

\begin{equation}
{\cal L}=\sum_{k}c_{k}F^{k}
\end{equation}

where the sum may involve both positive and negative powers of
$k$.\\

Here we consider the generalization of Maxwell's electro-magnetic
Lagrangian up to the second order terms of the fields as in the
form \cite{Camara2004}

\begin{equation}
{\cal L}=-\frac{1}{4\mu_{0}}~F+\omega F^{2}+\eta F^{*2}
\end{equation}

where, $\omega$ and $\eta$ are arbitrary constants. Here

\begin{equation}
F^{*}\equiv F^{*}_{\mu\nu}F^{\mu\nu}
\end{equation}

where, $F^{*}_{\mu\nu}$ is the dual of $F_{\mu\nu}$. Now, the
electro-magnetic field $F$ has the expression in terms of electric
field $E$ and magnetic field $B$ as in the form
$F=2(B^{2}-E^{2})$. So the corresponding energy-momentum tensor
for non-linear electro-magnetic theory has the form

\begin{equation}
T_{\mu\nu}=-4~\frac{\partial {\cal L}}{\partial
F}~F^{\alpha}_{\mu}F_{\alpha\nu}+\left(\frac{\partial {\cal
L}}{\partial F^{*}}~F^{*}-{\cal L}\right)g_{\mu\nu}
\end{equation}

Now we consider the homogeneous electric field $E$ in plasma gives
rise to an electric current of charged particles and then rapidly
decays. So the squared magnetic field $B^{2}$ dominates over
$E^{2}$, i.e., in this case, $F=2B^{2}$. So $F$ is now only the
function of magnetic field (vanishing electric component) and
hence the FRW universe may be called {\it Magnetic Universe}. Now
from equation (6), we obtain the expressions of magnetic density
and pressure as

\begin{equation}
\rho_{B} = \frac{B^{2}}{2\mu_{0}}(1-8\mu_{0}\omega B^{2})
\end{equation}
and
\begin{equation}
p_{B} =\frac{1}{6\mu_{0}}B^{2}(1-40\mu_{0}\omega B^{2})
=\frac{1}{3}\rho_{B}-\frac{16}{3}\omega B^{4}
\end{equation}

It is to be noted that the density $\rho_{B}$ of the magnetic
field must be positive, so the magnetic field $B$ must satisfy
$B<\frac{1}{2\sqrt{2\mu_{0}\omega}}$ with $\mu_{0}>0$ and
$\omega>0$. If
$\frac{1}{2\sqrt{6\mu_{0}\omega}}<B<\frac{1}{2\sqrt{2\mu_{0}\omega}}$,
the strong energy condition is violated i.e.,
$\rho_{B}+3p_{B}=\frac{B^{2}}{\mu_{0}}(1-24\mu_{0}\omega B^{2})<0$
and in this case, the magnetic field generates dark energy which
drives acceleration of the universe.\\

\section{\bf{Field Equations and Solutions of Higher Dimensional FRW Model}}

We consider the $(n+2)$-dimensional flat homogeneous and isotropic
universe described by FRW metric which is given by
\cite{Chatterjee, Utpal}
\begin{equation}
ds^{2}=-dt^{2}+a^{2}(t)[dr^{2}+r^{2}dx_{n}^{2}]
\end{equation}

where $a(t)$ is the scale factor and

\begin{equation}
dx_{n}^{2}=d\theta^{2}+sin^{2}\theta_{1}d\theta^{2}_{2}+......+
sin^{2}\theta_{1}sin^{2}\theta_{2}...sin^{2}\theta_{n-1}d\theta^{2}_{n}
\end{equation}

Now assume that the universe is filled with dark matter and
magnetic field type dark energy, so the Einstein's field equations
in higher dimension are given by

\begin{equation}
\frac{n(n+1)}{2}\left(\frac{\dot{a}}{a}\right)^{2}= \rho_{total}
\end{equation}
and
\begin{equation}
n\frac{\ddot{a}}{a}+\frac{n(n-1)}{2}\left(\frac{\dot{a}}{a}\right)^{2}=-p_{total}
\end{equation}
where

\begin{equation}
\rho_{total}=\rho_{m}+\rho_{B}=\rho_{m}+\frac{B^{2}}{2\mu_{0}}(1-8\mu_{0}\omega
B^{2})
\end{equation}

and

\begin{equation}
p_{total}=p_{m}+p_{B}=w_{m}\rho_{m}+\left(\frac{1}{3}\rho_{B}-\frac{16}{3}\omega
B^{4}\right)
\end{equation}

where $8\pi G=c=1$. Also $\rho_{m}$ and $p_{m}$ are the energy
density and pressure of the dark matter with the equation of state
given by $p_{m}=w_{m}\rho_{m}$, $-1\leq w_{m}\leq 1$ and
$\rho_{B}$ and $p_{B}$ are the energy density and pressure due to
magnetic field. The energy conservation equation in higher
dimensional cosmology is given by

\begin{equation}
\dot{\rho}_{total}+(n+1)H(\rho_{total}+p_{total})=0
\end{equation}

where $H$ is the Hubble parameter defined as
$H=\frac{\dot{a}}{a}$. According to the recent Supernovae and CMB
data, the energy transfer decay rate should be proportional to the
present value of the Hubble parameter. Then we consider the model
of interaction between dark matter and dark energy governed by the
magnetic field, through a phenomenological interaction term $Q$.
Hence the energy-conservation equation (15) becomes
\begin{equation}
\dot{\rho}_{m}+(n+1)H(\rho_{m}+p_{m})=+Q
\end{equation}
and
\begin{equation}
\dot{\rho}_{B}+(n+1)H(\rho_{B}+p_{B})=-Q
\end{equation}

For simplicity of the calculation, we take the interaction
component as \cite{Tanwi2011}

\begin{equation}
Q=(n+1)\delta\frac{B}{\mu_{0}}(1-16\mu_{0}\omega B^{2})H
\end{equation}

where, $\delta$ is a small positive quantity, termed as
interaction parameter. Using the above expressions of $\rho_{B}$,
$p_{B}$ and solving the equations (16)-(18) we obtain
\cite{Tanwi2011}

\begin{equation}
B=-\frac{3}{2}\delta+\frac{B_{0}+
\frac{3\delta}{2}}{(1+z)^{-\frac{2}{3}(n+1)}},~~~~~B_{0}~\text{being
a constant.}
\end{equation}

and

\begin{eqnarray*}
\rho_{m}=\frac{\rho_{m0}}{(1+z)^{-(n+1)(w_{m}+1)}}+
\frac{\delta}{2\mu_{0}(1+z)^{-(n+1)(w_{m}+1)}}\left[\frac{32\omega\mu_{0}(B_{0}+
\frac{3\delta}{2})^{3}}{w_{m}-1}\left(1-(1+z)^{-(n+1)(w_{m}-1)}\right)\right.
\end{eqnarray*}

\begin{eqnarray*}
\left.-\frac{3\delta(-1+36\delta^{2}\omega
\mu_{0})}{w_{m}+1}\left(1-(1+z)^{-(n+1)(w_{m}+1)}\right)-\frac{432\delta\omega\mu_{0}(B_{0}+
\frac{3\delta}{2})^{2}}{3w_{m}-1}\left(1-(1+z)^{-\frac{1}{3}(n+1)(3w_{m}-1)}\right)\right.
\end{eqnarray*}

\begin{equation}
\left.+\frac{6(B_{0}+
\frac{3\delta}{2})(-1+108\delta^{2}\omega\mu_{0})}{3w_{m}+1}
\left(1-(1+z)^{-\frac{1}{3}(n+1)(3w_{m}+1)}\right)\right]
\end{equation}

where $\rho_{m0}$ is a constant and redshift $z=\frac{1}{a}-1$.
From the above solutions, it may be concluded that the interaction
term $Q$ always decays with the evolution of the universe. If
$\delta= 0$, we get the non-interacting solutions, i.e., $B\propto
a^{-2\frac{(n+1)}{3}}$.  When $n=2$ (i.e., for 4D), we can recover
the result of Ref \cite{Tanwi2011} (i.e., $B\propto a^{-2}$) and
also when $\delta=0$, $n=2$ and we drop the matter density term
(i.e., $\rho_{m} = 0$), then we can verify the result of Ref.
\cite{Camara2004} for$\Lambda = 0$. Otherwise, our solutions are
not similar with their solutions.\\

The Hubble parameter and \textbf{Deceleration parameter} are given
by

\begin{equation}
H^{2}(z)=\frac{2\rho_{m}+\frac{B^{2}}{\mu_{0}}(1-8\mu_{0}\omega
B^{2})}{n(n+1)}
\end{equation}
and
\begin{equation}
q=-1-\frac{\dot{H}}{H^{2}}=\frac{n-1}{2}+\frac{n+1}{6}\left[
\frac{3w_{m} \rho_{m}-16 \omega
B^{4}+\frac{B^{2}}{2\mu_{0}}(1-8\mu_{0}\omega
B^{2})}{\rho_{m}+\frac{B^{2}}{2\mu_{0}}(1-8\mu_{0}\omega
B^{2})}\right]
\end{equation}

where $B$ and $\rho_{m}$ are given by (19) and (20). The
expression of $q$ is very complicated in terms of $z$. So the
variation of deceleration parameter $q$ against redshift $z$ is
plotted in figure 1 for 4D $(n=2)$, 5D $(n=3)$ and 6D $(n=3)$.
From figure, we see that $q$ decreases from some positive value to
$-1$ as $z$ decreases. So the model generates first deceleration
and then acceleration as universe expands. \\\\

$\bullet ${\normalsize\textbf{{\it Om} DIAGNOSTIC:}}\\

Recently, Sahni et al \cite{Sahni2003,Sahni2008} proposed a new
cosmological parameter named {\it Om} which is a combination of
Hubble parameter and the cosmological redshift and provides a null
test of dark energy. {\it Om} diagnostic has been discussed
together with statefinder for generalized Chaplygin gas model from
cosmic observations in \cite{Tong2009,Lu2009}. Generally, it was
introduced to differentiate $\Lambda$CDM from other dark energy
models. For $\Lambda$CDM model, ${\it Om} = \Omega_{m0}$ is a
constant, independent of redshift $z$. Also it helps to
distinguish the present matter density constant $\Omega_{m0}$ in
different models more effectively. The main utility for {\it Om}
diagnostic is that the quantity of {\it Om} can distinguish dark
energy models with less dependence on matter density
${\Omega_{m0}}$ relative to the EoS of dark energy. Our starting
point for {\it Om} diagnostics in the Hubble parameter and it is
defined as:

\begin{equation}
Om(z)=\frac{h^{2}(x)-1}{x^{3}-1}
\end{equation}
where $x = z+1$ and $h(x) = \frac{H(x)}{H_{0}}\equiv{\tilde{H}}$
and $H_{0}$ is the present value of the Hubble parameter. Now in
our interacting magnetic field  model, we obtain

\begin{equation}
Om(z)=\frac{\tilde{H}^{2}(z)-1}{(1+z)^{3}-1}=\frac{\frac{2\rho_{m}+\frac{B^{2}}{\mu_{0}}(1-8\mu_{0}\omega
B^{2})}{n(n+1)H_{0}^{2}}-1}{(1+z)^{3}-1}
\end{equation}

where $\tilde{H}^{2}(z)=\frac{H^{2}(z)}{H_{0}^{2}}$. We draw the
$Om$ diagnostic against redshift $z$ in figure 2 for 4D, 5D and
6D. The $Om$ diagnostic always increases as $z$ decreases (universe expands).\\

\begin{figure}
\includegraphics[scale=0.75]{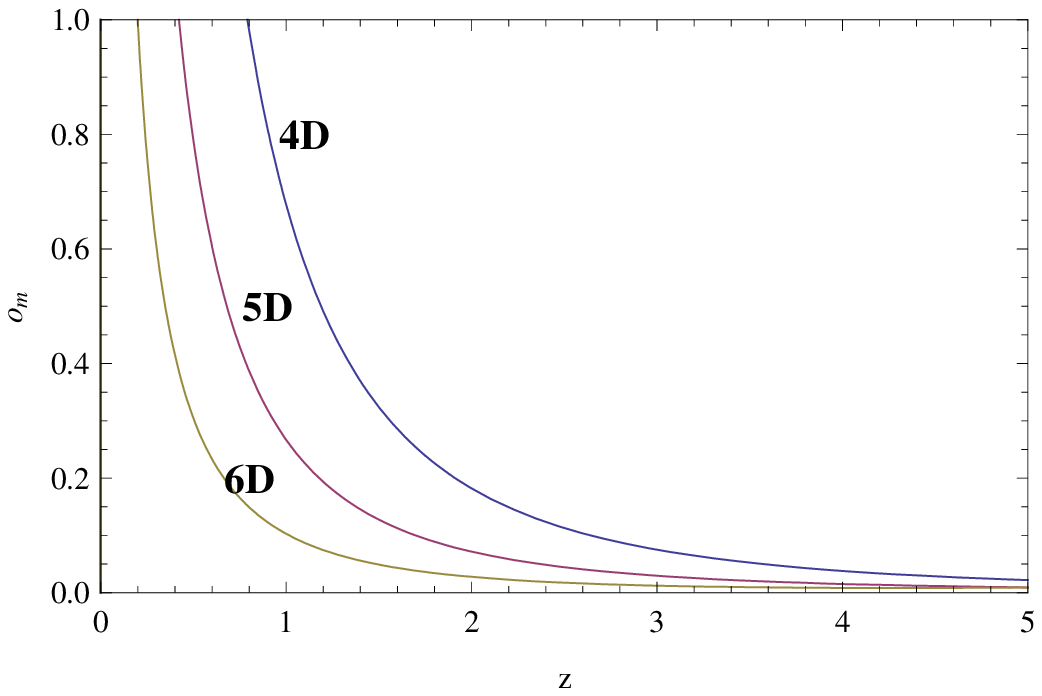}~~~~
\includegraphics[scale=0.75]{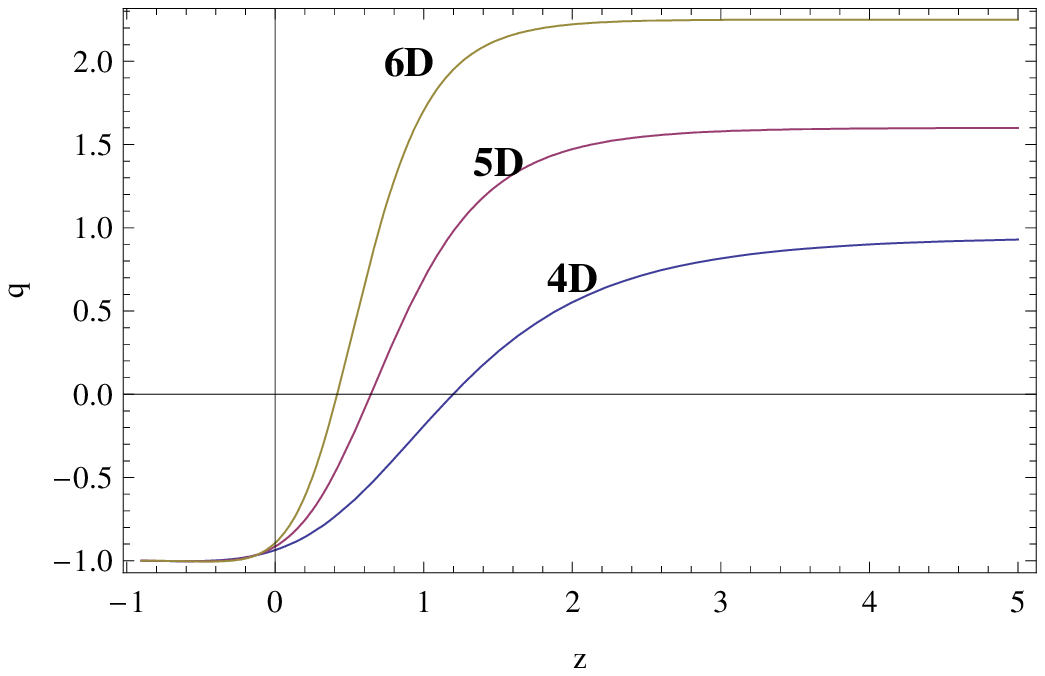}
\vspace{2mm}
~~~~~~Fig.1~~~~~~~~~~~~~~~~~~~~~~~~~~~~~~~~~~~~~~~~~~~~~~~~~~~~~~~~~~~~~~~~Fig.2~~\\
\vspace{4mm}

Figs. 1 and 2 represent the variations of deceleration parameter
$q$ and $Om$ diagnostic against redshift $z$ respectively for 4D,
5D and 6D.

\vspace{6mm}
\end{figure}

\section{\bf{Observational Constraints}}

In this section, we investigate the observational constraints of
the higher dimensional FRW model of the magnetic universe. We
shall determine the expected bounds of the theoretical parameters
by $\chi^{2}$ statistical best fit test with the basis of
$H(z)$-$z$ (Stern) \cite{Stern} and Stern+BAO \cite{Wu, Paul,
Paul1, Paul2, Paul3, Chak} joint data analysis. We also determine
the statistical confidence contours between two parameters
$\omega$ and $B_{0}$ in different dimensions. The best fit of
distance modulus for our theoretical model and the Supernova Type
Ia Union2 sample are analyzed for different dimensions. To
investigate the bounds of model parameters here we consider Stern
($H(z)$-$z$) data set with 12 data of $H(z)$-$z$ (Stern) in the
following Table 1 \cite{Stern}.

\[
\begin{tabular}{|c|c|c|}
\hline
  ~~~~~~~~~~~~$z$ ~~~~~~~~~~& ~~~~~~~~~~$H(z)$ ~~~~~~~~~~~~~& ~~~~~~~~~~~$\sigma(z)$~~~~~~~~~~~~\\
  \hline
  0 & 73 & $\pm$ 8 \\
  0.1 & 69 & $\pm$ 12 \\
  0.17 & 83 & $\pm$ 8 \\
  0.27 & 77 & $\pm$ 14 \\
  0.4 & 95 & $\pm$ 17.4\\
  0.48& 90 & $\pm$ 60 \\
  0.88 & 97 & $\pm$ 40.4 \\
  0.9 & 117 & $\pm$ 23 \\
  1.3 & 168 & $\pm$ 17.4\\
  1.43 & 177 & $\pm$ 18.2 \\
  1.53 & 140 & $\pm$ 14\\
  1.75 & 202 & $\pm$ 40.4 \\ \hline
\end{tabular}
\]
~~~~~~~~~~~~~~~~~~~~~~~~~~~~~~~~~~~~~~~~~{\bf Table 1:} $H(z)$ and
$\sigma(z)$ for different values of $z$.\\\\\\\\

Defining $\Omega_{m0}=\frac{\rho_{m0}}{\frac{n(n+1)}{2}H_{0}^{2}}$
and using the expression of $\rho_{m}$ and $B$, the expression of
$H^{2}(z)$ becomes

\begin{eqnarray*}
H^{2}(z)=\frac{H_{0}^{2}\Omega_{m0}}{(1+z)^{-(n+1)(w_{m}+1)}}+
\frac{\delta}{n(n+1)\mu_{0}(1+z)^{-(n+1)(w_{m}+1)}}\left[\frac{32\omega\mu_{0}(B_{0}+
\frac{3\delta}{2})^{3}}{w_{m}-1}\left(1-(1+z)^{-(n+1)(w_{m}-1)}\right)\right.
\end{eqnarray*}

\begin{eqnarray*}
\left.-\frac{3\delta(-1+36\delta^{2}\omega
\mu_{0})}{w_{m}+1}\left(1-(1+z)^{-(n+1)(w_{m}+1)}\right)-\frac{432\delta\omega\mu_{0}(B_{0}+
\frac{3\delta}{2})^{2}}{3w_{m}-1}\left(1-(1+z)^{-\frac{1}{3}(n+1)(3w_{m}-1)}\right)\right.
\end{eqnarray*}

\begin{eqnarray*}
\left.+\frac{6(B_{0}+
\frac{3\delta}{2})(-1+108\delta^{2}\omega\mu_{0})}{3w_{m}+1}
\left(1-(1+z)^{-\frac{1}{3}(n+1)(3w_{m}+1)}\right)\right]
\end{eqnarray*}

\begin{equation}
+\frac{\left(-\frac{3}{2}\delta+\frac{B_{0}+
\frac{3\delta}{2}}{(1+z)^{-\frac{2}{3}(n+1)}}\right)^{2}}{n(n+1)\mu_{0}}\left[1-8\mu_{0}\omega
\left(-\frac{3}{2}\delta+\frac{B_{0}+
\frac{3\delta}{2}}{(1+z)^{-\frac{2}{3}(n+1)}}\right)^{2}\right]
\end{equation}

This equation can be written in the form $H(z)=H_{0}E(z)$, where
$E(z)$ known as normalized Hubble parameter contains six model
parameters $\mu_{0},B_{0},\omega,\delta,n,w_{m_{}}$ beside the
redshift parameter $z$. Now to find the bounds of the parameters
and to draw the statistical confidence contour (66\%, 90\% and
99\% confidence levels) we fixed four parameters $\mu_{0},
n,\delta,w_{m}$. In the this case we find the bounds of
$B_{0},\omega$ and draw the contours between them. Here the
parameter $n$ determines the higher dimensions and we perform
comparative study between three
cases : 4D $(n=2)$, 5D $(n=3)$ and 6D $(n=4)$ respectively.\\

\begin{figure}
\includegraphics[scale=0.5]{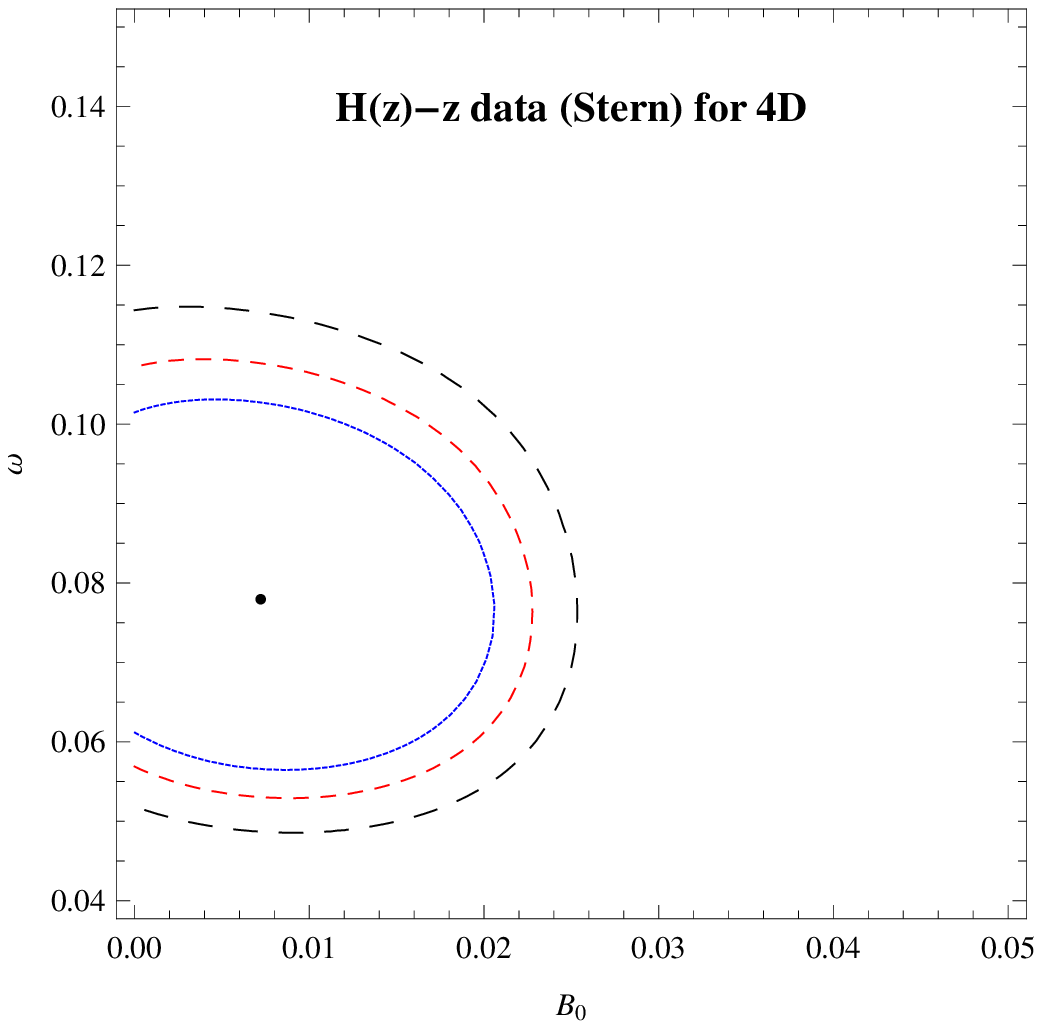}~~
\includegraphics[scale=0.5]{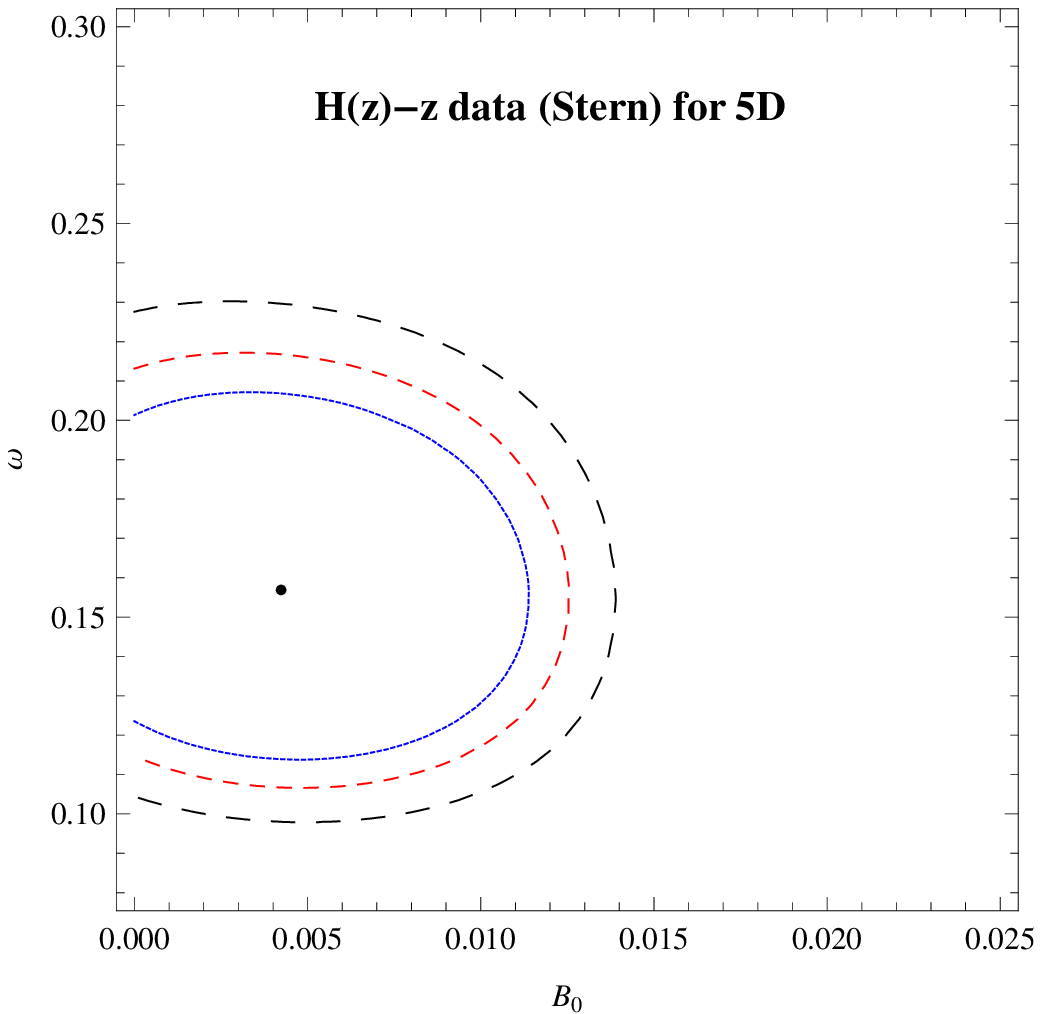}~~
\includegraphics[scale=0.5]{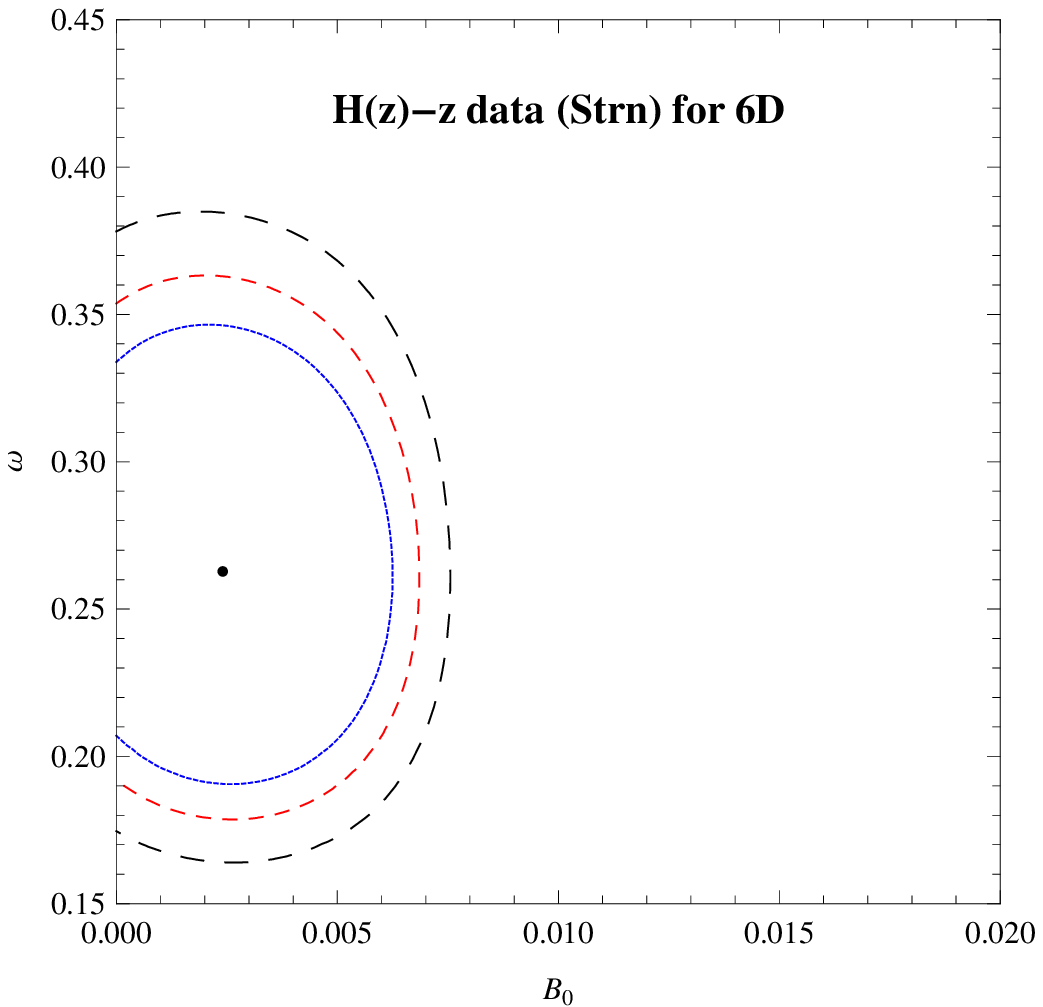}\\
\vspace{2mm}
~Fig.3~~~~~~~~~~~~~~~~~~~~~~~~~~~~~~~~~~~~~~~~~~~Fig.4~~~~~~~~~~~~~~~~~~~~~~~~~~~~~~~~~~~~~~~~~~~Fig.5\\
\vspace{4mm}

Figs. 3 - 5 show that the variations of $\omega$ against $B_{0}$
in 4D, 5D and 6D respectively for different confidence levels. The
66\% (solid, blue), 90\% (dashed, red) and 99\% (dashed, black)
contours are plotted in these figures for the $H(z)$-$z$ (Stern)
analysis.

\vspace{4mm}
\end{figure}

\subsection{Analysis with Stern ($H(z)$-$z$) Data Set}

Here we analyze the model parameters using twelve data
\cite{Stern} of  Hubble parameter for different redshift given by
Table 1. We first form the Chi square statistic (with 11 degree of
freedom) as a sum of standard normal distribution as follows:

\begin{equation}
{\chi}_{O}^{2}=\sum\frac{(H_{E}(H_{0},B_{0},\mu_{0},n,w_{m},\omega,\delta,
,z)-H_{OB})^{2}}{2\sigma^{2}}
\end{equation}

where $H_{E}$ and $H_{OB}$ are theoretical and observed values of
Hubble parameter at different redshifts respectively and $\sigma$
is the corresponding error. Here, $H_{0}$ is a nuisance parameter
and can be safely marginalized. We consider the observed
parameters $\Omega_{m0}=0.28$, $H_{0}$ = 72 $\pm$ 8 Kms$^{-1}$
Mpc$^{-1}$ and a fixed prior distribution. Here we shall determine
the model parameters $B_{0},\omega$ by minimizing the $\chi^{2}$
statistic. The reduced chi square can be written as

\begin{equation}
{\chi}_{R}^{2}=-2
\ln\int\left[e^{\frac{{\chi}_{O}^{2}}{2}}P(H_{0})\right]dH_{0}
\end{equation}

where $P(H_{0})$ is the prior distribution. We now plot the graphs
for different confidence levels (i.e., 66\%, 90\% and 99\%
confidence levels) and for three different dimensions (4D, 5D and
6D). Now our best fit analysis with Stern observational data
support the theoretical range of the parameters. When we fix the
parameters $\mu_{0}=0.7,\delta=0.01,w_{m}=0.1$, the 66\% (solid,
blue), 90\% (dashed, red) and 99\% (dashed, black) contours for
$(B_{0},\omega)$ are plotted in figures 3, 4 and 5 for 4D $(n=2)$,
5D $(n=3)$ and 6D $(n=4)$ respectively. The best fit values of
$(B_{0},\omega)$ and minimum values of $\chi^{2}$ for different
values of $n=2,3,4$ (i.e., different dimensions) are tabulated in
Table 2. For each dimension, we compare the model parameters
through the values of the parameters and by the statistical
contours. From this comparative study, one can understand the
convergence of theoretical values of the parameters to the values
of the parameters obtained from the observational data set and how
it changes from normal four dimension to higher
dimension (6D).\\

\[
\begin{tabular}{|c|c|c|c|}
\hline
  ~~~~~~$n$ ~~~~~& ~~~~~~~$B_{0}$ ~~~~~~~~& ~~~$\omega$~~~~~&~~~~~$\chi^{2}_{min}$~~~~~~\\
  \hline
  $~~2(4D)$ & 0.007 & 0.078 & 68.796 \\
  $~~3(5D)$ & 0.004 & 0.157 & 68.150 \\
  $~~4(6D)$ & 0.002 & 0.263 & 67.694 \\
   \hline
\end{tabular}
\]
{\bf Table 2:} $H(z)$-$z$ (Stern): The best fit values of $B_{0}$,
$\omega$ and the minimum values of $\chi^{2}$ for different
dimensions.

\begin{figure}
\includegraphics[scale=0.5]{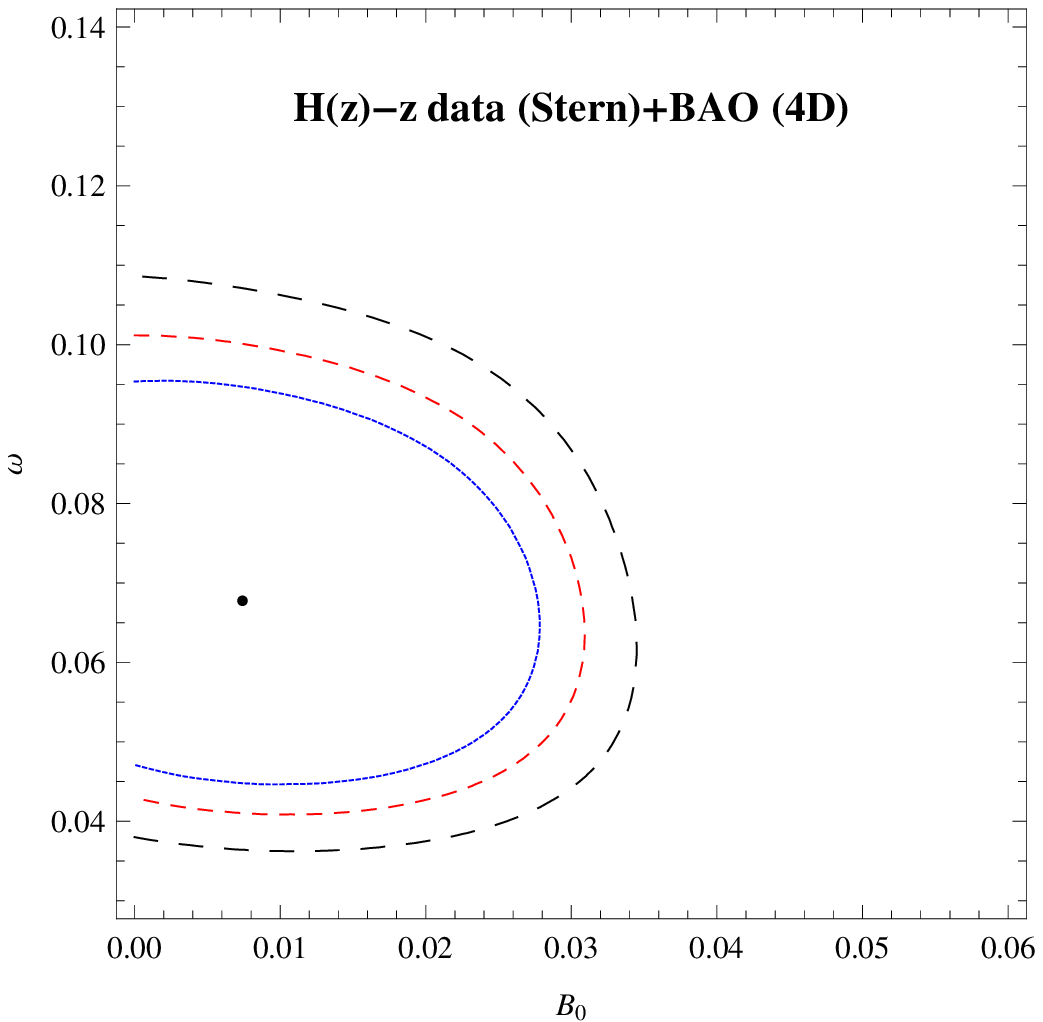}~~
\includegraphics[scale=0.5]{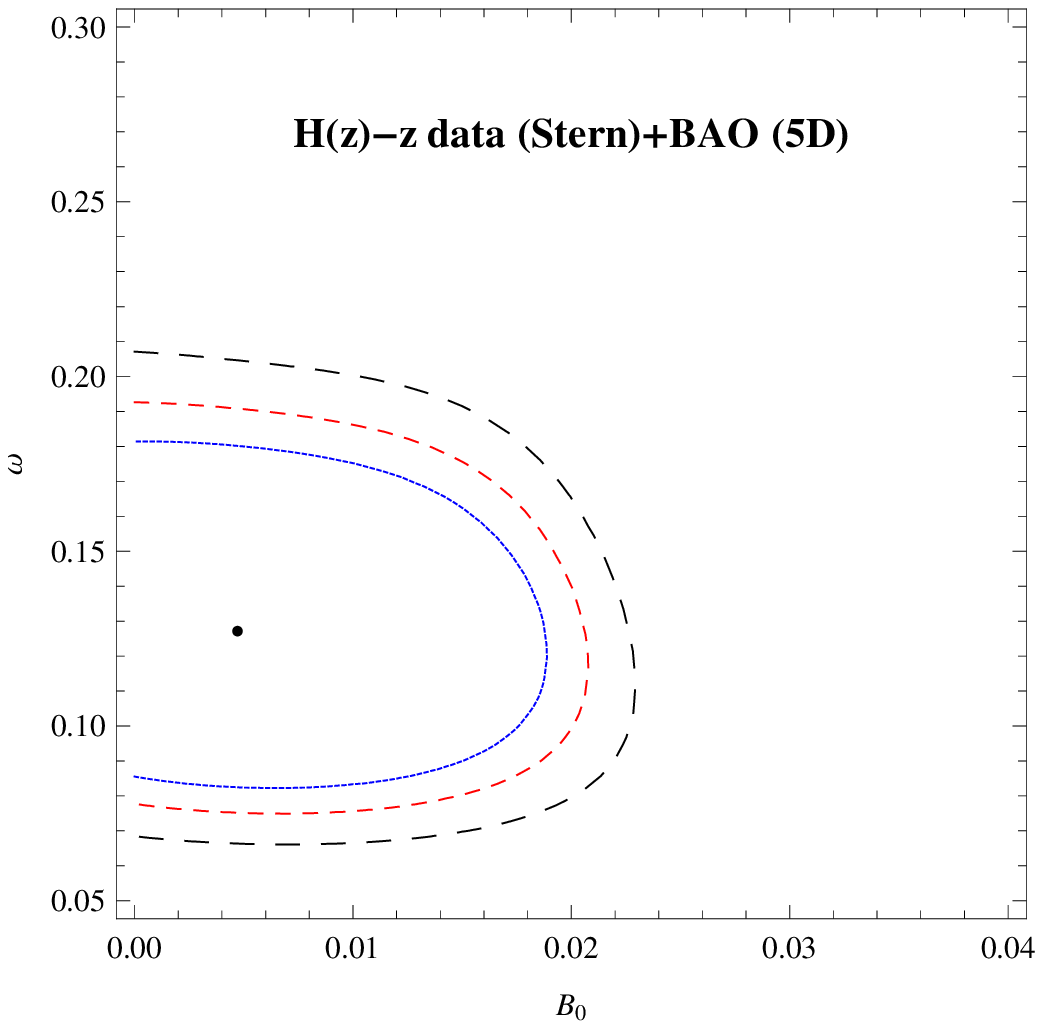}~~
\includegraphics[scale=0.5]{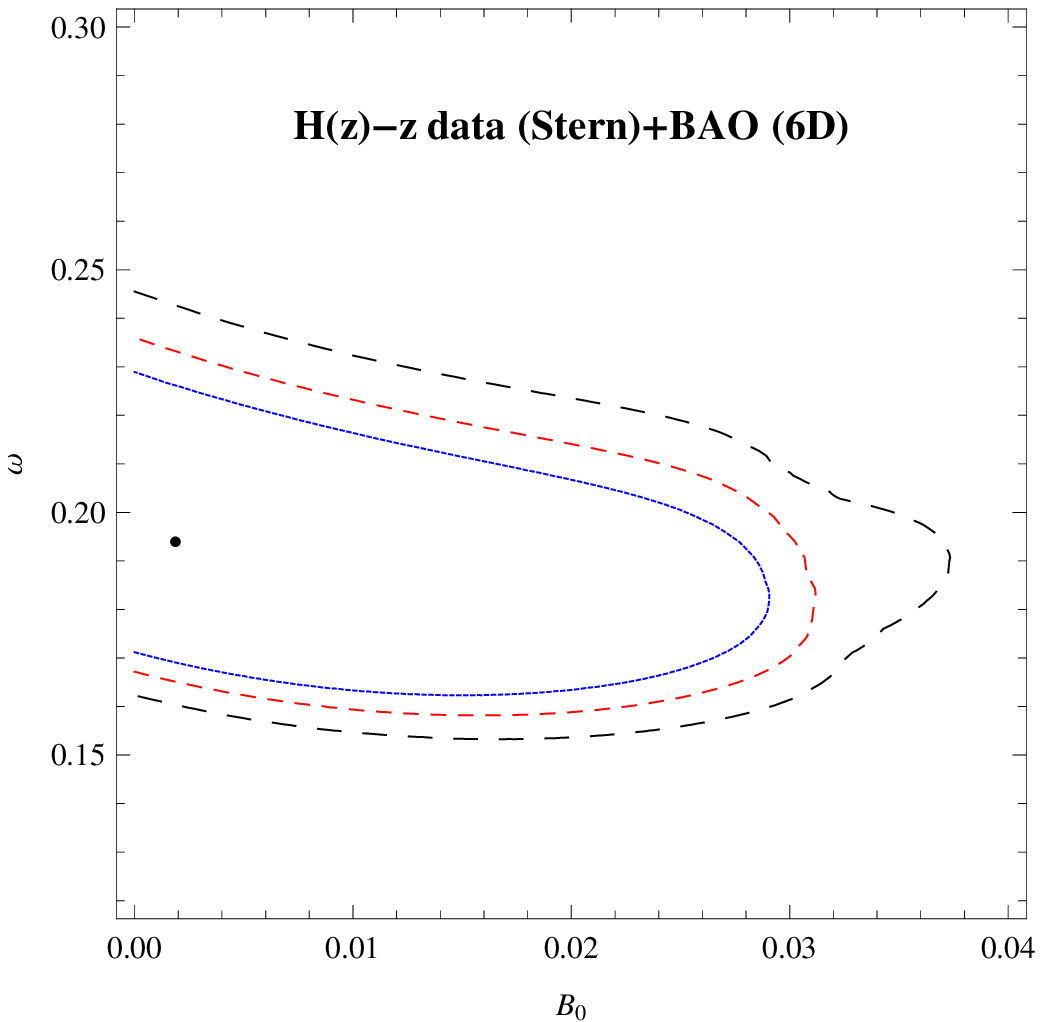}\\
\vspace{2mm}
~Fig.6~~~~~~~~~~~~~~~~~~~~~~~~~~~~~~~~~~~~~~~~~~~Fig.7~~~~~~~~~~~~~~~~~~~~~~~~~~~~~~~~~~~~~~~~~~~Fig.8\\
\vspace{4mm}

Figs. 6 - 8 show that the variations of $\omega$ against $B_{0}$
in 4D, 5D and 6D respectively for different confidence levels. The
66\% (solid, blue), 90\% (dashed, red) and 99\% (dashed, black)
contours are plotted in these figures for the $H(z)$-$z$
(Stern)+BAO joint analysis.

\vspace{4mm}
\end{figure}

\subsection{Joint Analysis with Stern $+$ BAO Data Sets}

The Baryon Acoustic Oscillations (BAO) in the primordial
baryon-photon fluid, leave a characteristic signal on the galaxy
correlation function, a bump at a scale $\sim$ 100 Mpc, as
observed by Eisenstein et al \cite{Eisenstein}. We shall
investigate the two parameters $B_{0}$ and $\omega$ for our model
using the BAO peak joint analysis for low redshift (with range $0
< z < 0.35$) using standard $\chi^2 $ distribution. The BAO peak
parameter may be defined by

\begin{eqnarray}
{\cal{A}}=\frac{\sqrt{\Omega_m}}{E(z_1)^{1/3}}\left(\frac{1}{z_1}~\int_{0}^{z_1}\frac{dz}{E(z)}\right)^{2/3}
\end{eqnarray}

where
\begin{eqnarray}
\Omega_m=\Omega_{m0}(1+z_1)^3 E(z_1)^{-2}
\end{eqnarray}

Here, $E(z)$ is the normalized Hubble parameter and $z_1 = 0.35$
is the typical redshift of the SDSS data sample. This quantity can
be used even for more general models which do not present a large
contribution of dark energy at early times. Now the $\chi^2$
function for the BAO measurement can be written as in the
following form

\begin{eqnarray}
\chi^2_{BAO}=\frac{({\cal{A}}-0.469)^2}{0.017^2}
\end{eqnarray}

where the value of the parameter ${\cal{A}}$ for the flat model
($k=0$) of the FRW universe is obtained by ${\cal{A}}=0.469 \pm
0.017$ using SDSS data set \cite{Eisenstein} from luminous red
galaxies survey. Now the total joint data analysis (Stern+BAO) for
the $\chi^2$ function defined by

\begin{eqnarray}
\chi^2_{Tot}=\chi^2_{Stern}+\chi^2_{BAO}
\end{eqnarray}

Now our best fit analysis with Stern$+$BAO observational data
support the theoretical range of the parameters. In figures 6-8,
we plot the graphs of $(B_{0},\omega)$ for different confidence
levels 66\% (solid, blue), 90\% (dashed, red) and 99\% (dashed,
black) contours for 4D, 5D and 6D respectively and by fixing the
other parameters $\mu_{0}=0.7,\delta=0.01,w_{m}=0.1$. The best fit
values of $(B_{0},\omega)$ and minimum values of $\chi^{2}$ for
different values of $n=2,3,4$ (i.e., different dimensions) are
tabulated in table 3.

\[
\begin{tabular}{|c|c|c|c|}
\hline
  ~~~~~~$n$ ~~~~~& ~~~~~~~$B_{0}$ ~~~~~~~~& ~~~$\omega$~~~~~&~~~~~$\chi^{2}_{min}$~~~~~~\\
  \hline
  $~~2(4D)$ & 0.007 & 0.068 & 790.463 \\
  $~~3(5D)$ & 0.005 & 0.127 & 777.620 \\
  $~~4(6D)$ & 0.002 & 0.194 & 595.719 \\
  \hline
\end{tabular}
\]
{\bf Table 3:} $H(z)$-$z$ (Stern)+BAO: The best fit values of
$B_{0}$, $\omega$ and the minimum values of $\chi^{2}$ for
different dimensions.

\subsection{Current Supernovae Type Ia Data}

In this section, we use Supernova Type Ia data at high redshifts
\cite{Perlmutter1998,Perlmutter1999,Riess1998,Riess2004} and
Baryonic Acoustic Oscillation (BAO) \cite{Eisenstein} to restrict
the parameters of the model for different dimensions. The
observations directly measure the distance modulus of a Supernovae
and its redshift $z$ \cite{Riess2,Kowalaski}. Now, take recent
observational data, including SNe Ia which consists of 557 data
points and belongs to the Union2 sample \cite{Amanullah}.\\

From the observations, the luminosity distance $d_{L}(z)$
determines the dark energy density and is defined by

\begin{equation}
d_{L}(z)=(1+z)H_{0}\int_{0}^{z}\frac{dz'}{H(z')}
\end{equation}

and the distance modulus (distance between absolute and apparent
luminosity of a distance object) for Supernovas is given by

\begin{equation}
\mu(z)=5\log_{10}
\left[\frac{d_{L}(z)/H_{0}}{1~\text{Mpc}}\right]+25
\end{equation}

The best fit of distance modulus as a function $\mu(z)$ of
redshift $z$ for our theoretical model and the Supernova Type Ia
Union2 sample are drawn for different dimensions (4D, 5D and 6D)
in figure 9 for our best fit values of $\omega$ and $B_{0}$ by
fixing the other parameters $\mu_{0}=0.7,\delta=0.01,w_{m}=0.1$.

\begin{figure}
\includegraphics[scale=0.8]{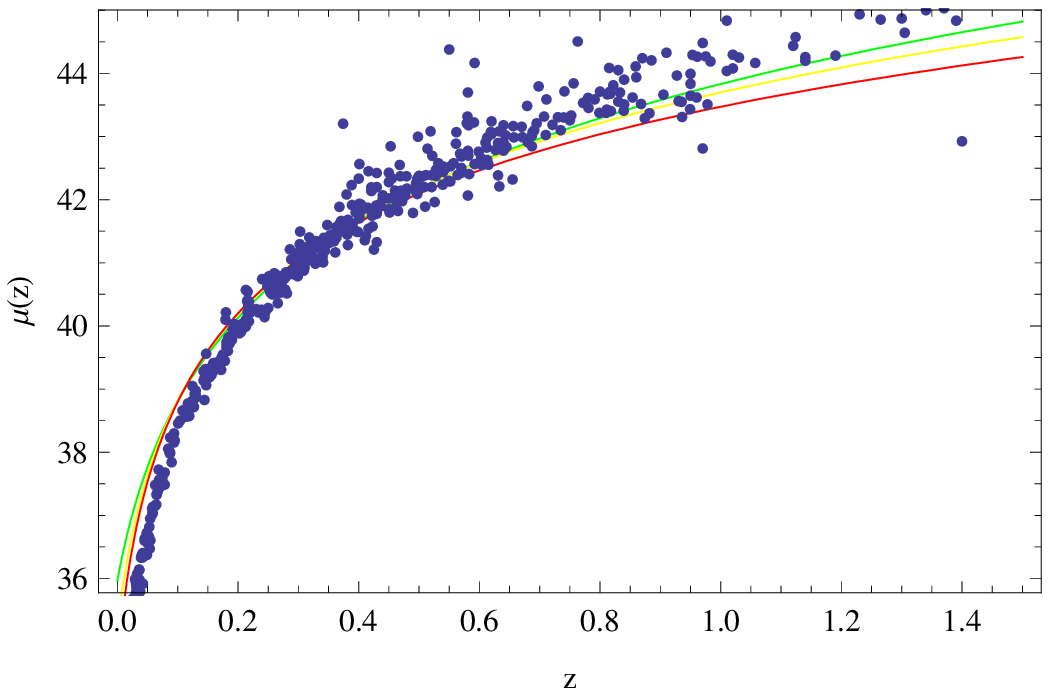}
\vspace{2mm}
~~~~~~~~~~~~~~~~~~~~~~~~~~~~~~~~~~~~~~~~~~~~~~~~~~~Fig.9~~~~~~~~~~~~~~~~~~~~~~~~~~~~~~~~~~~~~~~~~~~~~~~~~\\
\vspace{4mm}

In fig.9, $\mu(z)$ vs $z$ is plotted for our model (solid line)
(green for 4D, yellow for 5D and red for 6D) and the Union2 sample
(dotted points). \vspace{6mm}
\end{figure}

\section{\normalsize\bf{Discussions}}

In this work, we have considered the flat FRW model of the
universe in $(n+2)$-dimensions filled with the dark matter and the
magnetic field. We present the Hubble parameter in terms of the
observable parameters $\Omega_{m0}$ and $H_{0}$ with the redshift
$z$ and the other parameters like $B_{0},\omega,
\mu_{0},\delta,n,w_{m_{}}$. The magnetic field $B$ and
deceleration parameter $q$ have been calculated. The magnetic
field $B$ follows the power law form of redshift $z$. Now the
variation of deceleration parameter $q$ against redshift $z$ has
been plotted in figure 1 for 4D $(n=2)$, 5D $(n=3)$ and 6D
$(n=3)$. From figure, we see that $q$ decreases from some positive
value to $-1$ as $z$ decreases. So the model generates first
deceleration and then acceleration as universe expands. Recently
proposed $Om$ diagnostic has also been discussed for our model.
We draw the $Om$ diagnostic against redshift $z$ in figure 2 for
4D, 5D and 6D. The $Om$ diagnostic always increases as $z$ decreases
(universe expands).\\

We have investigated the observational constraints of the higher
dimensional FRW model of the magnetic universe. Here we have
chosen the observed values of $\Omega_{m0}=0.28$,
$\Omega_{x0}=0.72$ and $H_{0}$ = 72 Kms$^{-1}$ Mpc$^{-1}$. From
Stern data set (12 points), we have obtained the bounds of the
arbitrary parameters by minimizing the $\chi^{2}$ test. The
best-fit values of the parameters are obtained by 66\%, 90\% and
99\% confidence levels. Now to find the bounds of of the
parameters ($B_{0},\omega$) and to draw the statistical confidence
contour, we fixed four parameters
$\mu_{0}=0.7,\delta=0.01,w_{m}=0.1$ and $n=2,3,4$. Here the
parameter $n$ determines the higher dimensions and we perform
comparative study between three cases : 4D $(n=2)$, 5D $(n=3)$ and
6D $(n=4)$ respectively. We have plotted the graphs for different
confidence levels i.e., 66\%, 90\% and 99\% confidence levels and
for three different dimensions (4D, 5D and 6D) in figures 3-5. Now
our best fit analysis with Stern observational data support the
theoretical range of the parameters. The best fit values of
$(B_{0},\omega)$ and minimum values of $\chi^{2}$ for different
dimensions are tabulated in Table 2. For each dimension, we
compare the model parameters through the values of the parameters
and by the statistical contours. From this comparative study, one
can understand the convergence of theoretical values of the
parameters to the values of the parameters obtained from the
observational data set and how it changes from normal four
dimension to higher dimension (6D). Next due to joint analysis
with Stern+BAO observational data, we have also obtained the
bounds of the parameters $(B_{0},\omega)$ by fixing some other
parameters $\mu_{0}=0.7,\delta=0.01,w_{m}=0.1$ for 4D, 5D and 6D.
In figures 6-8, we have plotted the graphs of $(B_{0},\omega)$ for
different confidence levels 66\% (solid, blue), 90\% (dashed, red)
and 99\% (dashed, black) contours for 4D, 5D and 6D respectively.
The best fit values of $(B_{0},\omega)$ and minimum values of
$\chi^{2}$ for different dimensions are tabulated in Table 3. The
best fit of distance modulus as a function $\mu(z)$ of redshift
$z$ for our theoretical model and the Supernova Type Ia Union2
sample are drawn for different dimensions (4D, 5D and 6D) in
figure 9 for our best fit values of $\omega$ and $B_{0}$ by
fixing the other parameters $\mu_{0}=0.7,\delta=0.01,w_{m}=0.1$.\\\\\\\\

{\bf Acknowledgement:}\\

The authors are thankful to IUCAA, Pune, India for warm
hospitality where part of the work was carried out. Also UD is
thankful to CSIR, Govt. of India for providing research project
grant (No. 03(1206)/12/EMR-II).\\

\end{document}